\documentclass[prl,aps,floatfix, twocolumn]{revtex4} 

\usepackage{graphicx}
\usepackage{dcolumn}
\usepackage{bm}
\usepackage{longtable}
 \usepackage{amsmath}
 \usepackage{amssymb}
 \usepackage{eurosym}
\usepackage{float}
 
\begin{document}

\title{To bail-out or to bail-in? Answers from an agent-based model}

\author{Peter Klimek$^1$, Sebastian Poledna$^{1}$, J. Doyne Farmer$^{2,3}$, Stefan Thurner$^{1,3,4,}$}
\email{stefan.thurner@meduniwien.ac.at}

\affiliation{$^1$Section for Science of Complex Systems; Medical University of Vienna; 
Spitalgasse 23; A-1090; Austria\\ 
$^2$Institute for New Economic Thinking at the Oxford School and Mathematical Institute, Eagle House, Walton Well Rd., University of Oxford OX2 3ED; UK\\
$^3$Santa Fe Institute; 1399 Hyde Park Road; Santa Fe; NM 87501; USA\\
$^4$IIASA, Schlossplatz 1, A 2361 Laxenburg; Austria}

\begin{abstract} 
Since beginning of the 2008 financial crisis almost half a trillion euros have been spent to financially assist EU member states in taxpayer-funded bail-outs.
These crisis resolutions are often accompanied by austerity programs causing political and social friction on both domestic and international levels.
The question of how to resolve failing financial institutions under which economic preconditions is therefore a pressing and controversial issue of vast political importance.
In this work we employ an agent-based model to study the economic and financial ramifications of three highly relevant crisis resolution mechanisms.
To establish the validity of the model we show that it reproduces a series of key stylized facts if the financial and real economy.
The distressed institution can either be closed via a purchase \& assumption transaction, it can be bailed-out using taxpayer money, or it may be bailed-in in a debt-to-equity conversion.
We find that for an economy characterized by low unemployment and high productivity the optimal crisis resolution with respect to financial stability and economic productivity is to close the distressed institution.
For economies in recession with high unemployment the bail-in tool provides the most efficient crisis resolution mechanism.
Under no circumstances do taxpayer-funded bail-out schemes outperform bail-ins with private sector involvement.
 \end{abstract}
 
\maketitle

\section{Introduction}

In March 2013 Cyprus became the epicenter of financial turmoil in what would become the 2012-2013 Cypriot financial crisis.
The crisis gained momentum in the wake of the Greek government-debt crisis, when Cypriot banks were exposed to haircuts of up to 50\% in 2011 and the state was unable to raise liquidity from the markets to support its financial sector \cite{Stavarek13}.
In March 2013 bonds issued by Cyprus were downgraded to Junk status, which disqualified them from being accepted as collateral at the European Central Bank \cite{Wilson12}.
Consequently the Cypriot government requested financial aid from the European Financial Stability Facility (EFSF) \cite{AlJazeera13}.
What followed was an unprecedented, international struggle about who has to pay for the losses incurred by a national banking crisis.
The EU and the International Monetary Fund (IMF) replied to Cyprus' request by proposing a \euro10 billion deal, including a 6.7\% one-time bank deposit levy for deposits up to \euro100,000, and 9.9\% for higher deposits on all national bank accounts \cite{Stavarek13}.
Despite being on the verge of financial collapse, large demonstrations and political upheaval led  the Cypriot parliament to reject this proposal \cite{Guardian13}.
The situation was finally resolved by approving a plan to restructure the second largest Cypriot bank into a bad bank and to guarantee all deposits below \euro100,000, but to levy all higher uninsured deposits \cite{Stavarek13}.

In the case of Cyprus the EU-IMF originally proposed to resolve the banking crisis by a {\it bail-out}, i.e. by forcing {\it all} taxpayers or depositors in the country to participate in the \euro10 billion loan.
As a result of political turmoil, this plan was altered such that only depositors at the failing bank were forced to participate in the deal.
This took the form of a balance sheet restructuring, a debt-to-equity conversion.
This type of crisis resolution was the first realization of a so-called {\it bail-in} \cite{Otker11}.
The 'bail-out versus bail-in' debate, i.e. which crisis resolution outperforms the other in terms of fostering financial stability and overall economic output and growth, has been in full swing since then \cite{Zhou12, DeGrauwe13}.
Proponents of bail-ins often cite the moral hazard problem of bail-outs, i.e. the incentive to take risks for systemically important financial institutions (SIFI) when others (taxpayers) will have to pay eventual losses \cite{Zhou12}.
On the other hand, bail-ins are criticized for providing a channel for contagion risks from the failing institution to its investors, as has been pointed out in the discussion surrounding the resolution of the Austrian Hypo Alpe-Adria bank \cite{Reuters13}.
After all, it was a debt restructuring in the Greek banking sector similar to a bail-in which triggered the crisis in Cyprus \cite{Stavarek13}.
It is finally also unclear under which circumstances an orderly liquidation of a troubled bank would be preferable over both, bail-outs and bail-ins \cite{Stern04}.

The Cypriot financial crisis is only one in a series of examples for a banking crisis which blurred the lines between bank bail-outs and sovereign bail-outs.
For example, following the Emergency Economic Stabilization Act of 2008, the US treasury department disbursed loans of a combined volume of \$204.9 billion (b) among 707 banks in the Capital Purchase Program, of which \$16.7b are still outstanding \cite{GAO12}.
In Europe, the EFSF disbursed loans of \euro52b to Portugal \euro41b to Ireland in 2011 \cite{EFSF13, EC13}.
The first and second Greek bail-out consisted of tranches of about \euro110b each.
Spain received a loan of  \euro41.4b in 2012-2013.
In addition, the IMF also provided billions of loans as financial aid to European nations, \euro48.1b to Greece, \euro9.1b to Hungary, \euro22.5b to Ireland, \euro26b to Portugal, and \euro12.6b to Romania.
As a result of the 2008 financial crisis the combined bail-out volume in EU member states totals almost half a trillion Euro \cite{EFSF13, EC13}.
These bail-outs are typically accompanied by strict austerity programs causing political and social friction on both domestic and international levels.
The question of how to resolve a failing SIFI is ultimately a question about maintaining financial, economic and political stability on a supranational scale.
The performance of financial crisis resolution mechanisms has to be evaluated not only by ensuring financial stability, but also by how they impact the entire economy in terms of unemployment, economic growth, liquidity provision to entrepreneurs, etc.

DSGE models, by now the most popular way to study effects of policy interventions on macro-economic fluctuations, generally lack interactions between the financial sector and the real economy.
There are basically three approaches to introduce financial friction or defaults in these models.
First, in the collateral approach it is assumed that borrowers are required to provide a sufficient amount of collateral to guarantee that under no conditions in the future the borrower will have to default -- a cash-in-advance constraint \cite{Kiyotaki97}.
In the second approach banks are assumed to be able to hedge credit risk such that no defaults are possible at the expense of an external financing premium \cite{Bernanke99}.
In both approaches default does not occur in equilibrium, however.
In a third approach agents are allowed to choose what fraction of their outstanding debt to repay, and partially default on their obligations in return for a default penalty \cite{Dubey05}.
This approach introduces endogenous default rates at equilibrium.
By studying general equilibrium (GE) models which incorporate the cash-in-advance constraint and endogenous default rates, it has been shown that capital requirements for banks effectively lead to a trade-off between financial stability and economic efficiency \cite{Tsomocos03, Goodhart06, DeWalque10}.
A GE model where households take loans from both a banking system and a 'shadow banking system' was recently proposed in \cite{Goodhart12}.
There it was shown that if households choose to default on their loans, this may trigger forced selling by the shadow banks and lead to a fire sale dynamics.

In this work we address the question of which crisis resolution mechanisms perform optimally under given economic circumstances.
We focus on three highly relevant resolution mechanisms to contribute to both the 'bail-in versus bail-out', and the 'too-big-to-fail' debates.
(i) The troubled financial institution or bank is {\it liquidated by a purchase \& assumption} (P\&A) operation \cite{McGuire12}.
(ii) The bank is {\it bailed-out} using taxpayer money, or (iii) the distressed bank is {\it bailed-in} through a debt-to-equity conversion.
In particular try to clarify which crisis resolution mechanisms minimize financial contagion risks, lead to the highest liquidity provision for the economic sector, reduce unemployment most, lead to the highest economic output.
The first, P\&A resolution mechanism corresponds to the case where the distressed financial institution defaults and is closed down.
In the bail-out and bail-in cases the distressed bank continues to operate after the crisis.
This allows to investigate whether it is beneficial to let the failing institution default or not.
If the troubled bank is saved, we can inquire on a quantitative basis whether private sector involvement in the resolution plan is beneficial or not.
We employ the framework of the Mark I CRISIS model developed within the CRISIS project \footnote{http://www.crisis-economics.eu}.
This is an agent-based model (ABM) framework implementing a closed economy consisting of banks, firms, and households.
These agents interact with each other in various markets.
Within this framework and closely related models it has already been studied how systemic risk can be reduced or eliminated via a systemic risk transaction tax, or under different regulatory regimes \cite{Thurner13, Poledna14}.
These works also emphasized the importance of cascading spreading of credit risk within financial or interbank networks, and how this may lead to failures of the entire financial system \cite{Boss05, Iori08, Cacciolo12, Battiston12, Thurner12, Poledna12}.
The CRISIS framework is especially useful for the study of different crisis resolution mechanisms since it provides a model for both the economic and the financial sector.
Both sectors are modeled with fine granularity and a rich structure of inter-sector linkages, such as credit or deposit markets \cite{DelliGatti08, Gaffeo08, DelliGatti11}.
This makes it an ideal testbed to study the propagation of financial crises to the real economy and vice versa, and how this is influenced by different crisis resolution mechanisms.

In this work we address only mechanical aspects of bank defaults and resolutions.
There are no agents that can adapt their behavior to the given regulatory regime.
For instance, we do not include the moral hazard problem, i.e. that the knowledge of being bailed-out in financial distress induces additional risk-taking behavior by banks.
We also do not address the case of mixed crisis resolution strategies, such as letting banks default partially.
Each bank in the model follows the same strategy.
Hence there is also no long-term growth in the model generated by evolutionary culling banks with bad strategies, as envisaged in Schumpeter's concept of creative destruction \cite{Schumpeter42, Klimek12}.

\section{Crisis resolution mechanisms}
 
We focus on three highly relevant crisis resolution mechanisms in their generic forms.
One of them, 'purchase \& assumption' provides an orderly way to close a defaulting institution.
This will be compared to bail-out and bail-in resolution mechanisms, respectively, both ensuring a survival of the bank in distress.
We assume that the resolution mechanisms will be enacted 'top-down' by a single resolution authority acting within a legal framework which grants the required powers to the authority.
Here we provide the rationale behind each crisis resolution mechanism, their detailed model implementations follow in section \ref{sec:abm}.

{\bf Purchase \& Assumption.}
A 'purchase \& assumption' (P\&A) is a resolution mechanism which allows for transferring the troubled bank's operations to other, healthy banks \cite{McGuire12, FDIC03}.
The mechanism typically includes the withdrawal or cancellation of the troubled bank's license.
 Each of the other banks in the system purchases parts of the failing bank's assets and assumes its liabilities.
Here we assume that the volume of the asset purchase for each bank is proportional to the value of the purchasing bank's liquid assets.
Similarly, the assumption of the troubled bank's liabilities are proportional to the assets taken over.
Note that there exists a finance gap since the total asset values will be smaller than the combined value of the liabilities.
This gap will be closed by the banks which take over the troubled banks, since they get more liabilities than assets in the procedure described above.
Therefore the losses of the failing bank will effectively be paid by the other banks.
The main difference between a liquidation and a P\&A is that under liquidation the assets of a liquidated institution are sold over time to pay its liabilities to depositors,
whereas in the P\&A assets and liabilities are transferred to other banks.
An advantage of a P\&A is that there is no need to impose a process to pay out depositors of a failing bank.
Since the deposits are transferred to a healthy institution depositors may access their accounts without delay at each stage of the resolution.
Due to these reasons a P\&A is typically considered more efficient than a liquidation \cite{McGuire12}, it will be used here as the standard way to close down an institution.

{\bf Bail-out.}
A bail-out usually describes giving a loan to a financially distressed institution or country which is deemed healthy enough to survive after  recapitalization \cite{Wright09}.
Another reason to bail out an institution may be to minimize contagion risks of the insolvency of a large and interconnected, i.e. systemically important, financial institution.
We are interested in the case where the loan is not provided by a private investor, for example by buying floundering stocks of a company at firesale-prices, but by a government at the expense of taxpayers' interests.
In exchange for providing funds, the government typically receives preferred stock and therefore cash dividends over time, which are used to protect the taxpayers' money.
The government effectively becomes the owner of the taken-over institution whose common stock equity will be canceled (i.e. shareholders lose their investment), but the claims of debtors and depositors will be protected.
The use of bail-out resolution mechanisms has been highly controversial.
The existence of government-sponsored safety nets may actually work as an incentive for institutions to take financial risks, since in case of failure they will be bailed out anyhow -- the problem of moral hazard \cite{Zhou12}.
Another criticism involves the high costs typically involved in bail-outs \cite{Zhou12}.
In a sample of bail-outs in 40 different countries the average cost of a typical bank bail-out was estimated to be 12.8\% of GDP \cite{Wright09}.
Especially after the 2008 financial crisis, interest in alternative and cheaper crisis resolution mechanisms which do not require taxpayer involvement surged.

{\bf Bail-in.}
As opposed to a bail-out, a bail-in forces the creditors of the troubled financial institutions to bear some of the financial burden \cite{Otker11, Zhou12, DeGrauwe13}.
The claims of typically unsecured debt holders are written off in a bail-in and/or converted into equity to recapitalize the failing institution.
The bail-in instrument therefore provides a private sector funded resolution mechanism as opposed to government-funded solutions like bail-outs.
The legal framework for such top-down balance sheet restructuring was provided in the US by the Dodd-Frank Wall Street Reform in 2010, and in UK under the 2009 Banking Act \cite{FDIC12}.
In the Eurozone there is currently no legal framework for bail-ins.
The first realization of a bail-in took place in the Cypriot banking crisis 2012/2013.
Consequently there is relatively little experience in the long-time consequences of this resolution mechanism.
However, the perceived success of the Cyprus episode led the head of the Eurogroup of finance ministers, Jeroen Dijsselbloem, to state that the Cyprus deal may serve as a template for future crisis resolutions \cite{Stavarek13}.

\section{Agent-based model}
\label{sec:abm}
We study the performance of the crisis resolution mechanisms using the Mark I CRISIS model.
This is one of a suite of models developed within the CRISIS project on the basis of an agent-based macro-economic model \cite{DelliGatti08, Gaffeo08, DelliGatti11}.
The Mark I CRISIS model consists of a coupled economic and financial ABM which is {\it closed}, i.e. there are no in-flows and out-flows of any kind of capital.
Banking crises can thus not be resolved by simply printing new money -- some agents have to actually pay for the losses.
A similar version of the Mark I CRISIS framework has recently been used to study the implementation of a taxation scheme for interbank transactions in order to eliminate or reduce systemic risk \cite{Poledna14}.
For a more comprehensive description of the coupled economic-financial simulator see \cite{Poledna14}.
In the following we provide an overview of the agents and their interaction mechanisms, and list the changes with respect to previous versions of the CRISIS model.
The model contains three types of agents: households, banks, and firms.
Agents interact on various markets.
Households and firms interact on the job and consumption-good market, banks and firms interact on the credit market, banks interact on the interbank market.

{\bf Households.}
There are two types of household agents, namely $I$ firm owners and $J$ workers.
Each worker applies for a job at $z$ different firms in one model time-step.
Once hired, he/she receives a fixed income $w$ per time-step for supplying fixed labor productivity $\alpha$.
The workers deposit their income at one of the $B$ banks.
The banks where a given worker opens his/her deposit are randomly chosen in the model initialization and remain fixed throughout time (unless the bank is closed under a P\&A).
Each worker $j$ has a personal account $PA_{j,b}(t)$ at bank $b$.
Each of the $I$ firm owners owns exactly one firm, and both the firms and their owner are indexed by $i$.
They also have a personal account $PA_{i,b}(t)$ at a randomly chosen bank $b$.
At each time step each household -- worker or firm owner -- computes its consumption budget as a fixed percentage $c$ of its personal account.
It spends $c PA_{i/j,b}(t)$ on the cheapest single product it finds by comparing products from $z$ randomly chosen firms.

{\bf Firms.}
There are $I$ firms, all producing a perfectly substitutable good.
Each firm is owned by a single firm owner, and each firm owner always owns not more than one firm.
At each time step the firms have to decide on two quantities, their expected demand $d_i(t)$, and their expected price for the produced product, $p_i(t)$.
Let $\bar p(t)$ denote the average weighted price over all products, $\bar p(t) = \tfrac{\sum_i p_i(t) d_i(t)}{\sum_i d_i(t)} $.
They compute the demand and price by taking into account the previous demand $d_i(t-1)$ and price $p_i(t-1)$.
If the firm sold all of its goods at the previous time step, it either increases the price (if $p_i(t-1) < \bar p(t-1)$), or increases its expected demand if $p_i(t-1) \geq \bar p(t-1)$.
If the demand at the previous time step was lower than expected, the firm either reduces the price (if $p_i(t-1) > \bar p(t-1)$), or decreases the expected demand if $p_i(t-1) \leq \bar p(t-1)$.
In all other cases the expected demand and price do not change.
Each firm computes the number of required workers to achieve the desired demand, always assuming that each worker supplies labor productivity $\alpha$.
If the wages for this workforce exceed the firm's current liquidity, it applies for a credit.
On the credit market firms approach $z$ randomly chosen banks and choose the credit offered at the lowest rate.
If the real interest rate exceeds a threshold rate $r^{max}$, the firm's credit demand contracts to $\phi$ percent of the original loan volume.
After the loan is provided, firms recompute their desired workforce and hire or fire workers in order to meet the expected demand.
The newly produced goods are then sold on the consumption goods market.
A firm with positive profit pays out $\delta$ percent of the profit as dividend to its owners.
If a firm has negative liquidity after the consumption good market closes, the firm owner may provide the missing cash from his/her personal account.
If these funds are smaller than the liquidity gap, the firm goes bankrupt.
The firm owner is held liable and his/her personal account is used to partly pay off debtors (banks), which incur a capital loss in proportion to their investment.
Finally, the firm owner immediately starts a new company with $d_i(t) = \langle d_i(t) \rangle_i$ and $p_i(t) = \langle p_i(t) \rangle_i$ being equal to their current population average.

{\bf Banks.}
Banks are subject to capital requirements, their leverage must not exceed a given value $\lambda^{max}$; they have to maintain a cash reserve ratio of at least $\kappa^{min}$.
Each bank $b$ offers each firm $i$ a loan at rate $r_{b,i}$ based on $i$'s  leverage $l_i(t)$ (the quotient of $i$'s outstanding debt and its cash), 
\begin{equation}
r_{b,i} = r_0 (1+\epsilon) \left[ 1+\tanh(\mu l_i(t)) \right] \quad,
\label{interestrate}
\end{equation}
where $r_0$ is the refinancing rate, $\mu$ is a constant, and $\epsilon$ is a random number drawn from a uniform distribution between zero and one.
The nominal market interest rate $i^{(n)}(t)$ in the model is given by $i^{(n)}(t) = \tfrac{\sum_{i,b} r_{b,i}L{b,i}}{\sum_{i,b} L_{b,i}}$, where $L_{b,i}$ denotes the loan volume provided from bank $b$ to firm $i$ at rate $r_{b,i}$.
$CV(t)$ denotes the total amount of outstanding loans from all banks at a given time $t$.
As long as the banks have enough liquidity and fulfill their capital requirements, they always grant the requested loans. 
If they do not have enough cash, they approach $z$ randomly chosen other banks and try to get the missing amount from them.
Banks always grant interbank loans at a rate $r^{(ib)}$ when requested, given that they have enough cash and fulfill the capital requirements.
If a bank does not have enough cash and is unable to get the required funds from other banks, it does not pay out the loan.
Each time step firms reimburse $\tau$ percent of their outstanding debt.
In case the banks make a profit, they pay out $\delta$ percent of the profit as dividends.
If a bank's equity becomes negative it is insolvent and undergoes one of three possible crisis resolution mechanisms, (i) P\&A, (ii) bail-out, (iii) or bail-in.

\begin{figure*}
\begin{center}
 \includegraphics[width=140mm]{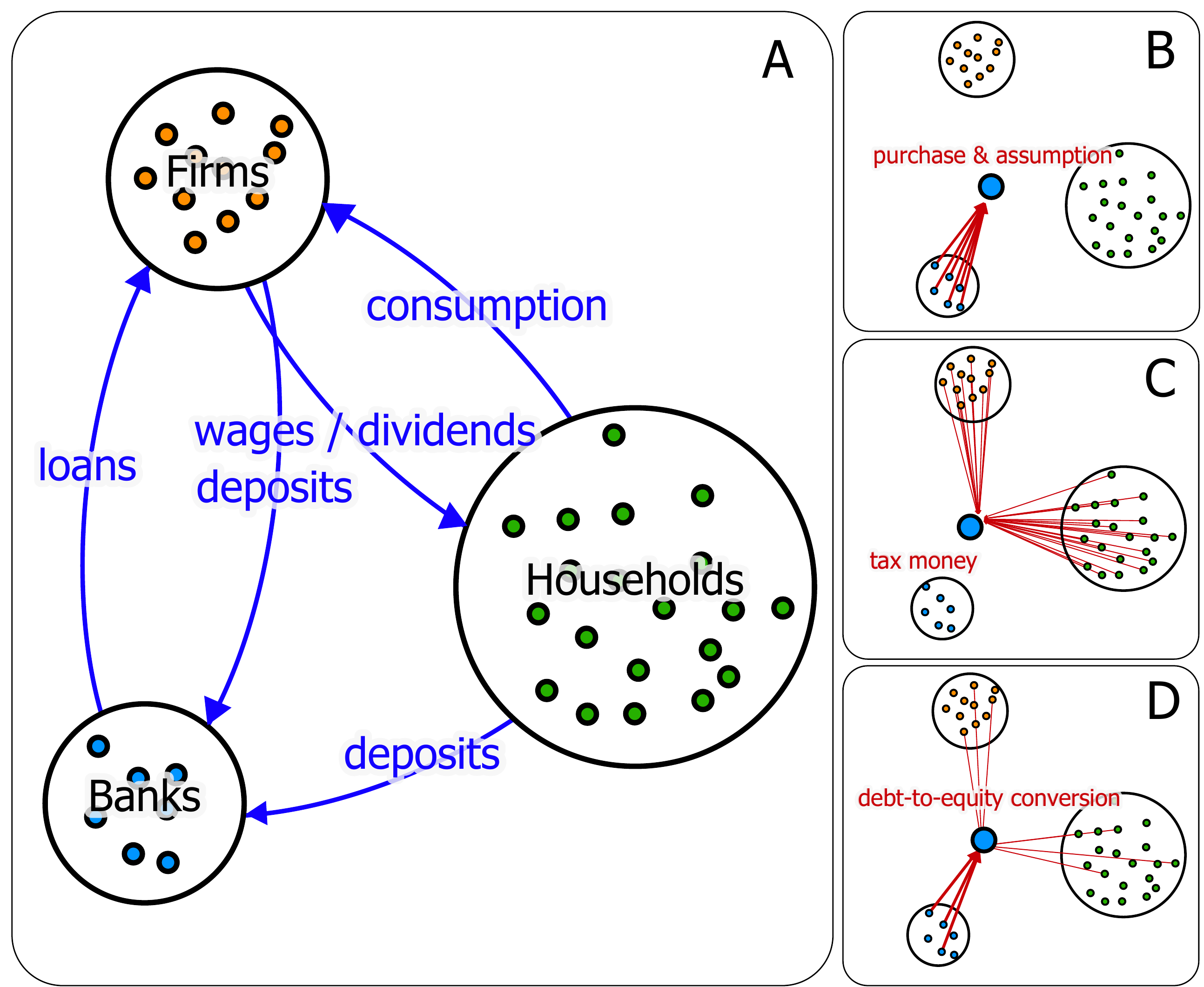}
\end{center}
 \caption{(A) The three sets of agents (households, firms, and banks) and their most important interactions are shown. Firms and households interact on the consumer-goods and job market, firms and banks on the credit market. Households and firms deposit their income with banks. The flows of funds in the CRISIS macro-financial model are shown in blue. Panels (B-D) show the flows of funds taking place in crisis resolutions as red arrows. The troubled bank $b$ is positioned in the middle of each panel. (B) Healthy banks purchase $b$'s assets and assume its liabilities. (C) Households and firms bail out $b$ using tax money. (D) A bail-in with a debt-to-equity conversion with $b$'s debtors takes place. }
 \label{Model}
\end{figure*}

The model interactions are summarized in figure \ref{Model}A.
Firms pay households wages or dividends, households consume goods produced by the firms.
Both firms and households make deposits at the banks.
The banks grant loans to the firms.
In each time-step the model consists of the following sequence of steps.
\begin{enumerate}
\item Firms define their labor demand and seek loans from banks.
\item Banks raise liquidity on the interbank market to service the loans to firms.
\item Firms hire or fire workers and produce goods.
\item Workers receive wages, spend their consumption budget on goods and save the remainder.
\item Firms pay out dividends, firms with negative cash go bankrupt.
\item Banks and firms repay loans.
\item Illiquid banks seek funds on the interbank market, insolvent banks are resolved.
\end{enumerate}

{\bf Crisis resolution mechanisms in the model.}
Bank $b$ goes bankrupt if its equity $E_b(t)$ becomes negative by an amount of $-M_b$, $E_b(t) \equiv -M_b < 0$.
The three different crisis resolution mechanisms are implemented in the following way.
\begin{description}
\item[P\&A] The normalized weight vector $w_i, \sum_i w_i = 1$ is constructed where the index $i$ runs over all banks with positive equities, i.e. healthy banks.
The entries in $w_i$ are proportional to the banks' equities.
Each healthy bank takes over a share proportional to $w_i$ of the resolved bank's interbank loans and firm loans.
The household deposits are also transferred from the failing to the healthy banks.
Each deposit is transferred to bank $i$ with probability $w_i$.
The flows of funds involved in the P\&A are sketched in figure \ref{Model}B.
\item[Bail-out] The normalized weight vector $w_i$ runs over all households and firms and has entries proportional to the households' personal accounts $PA_{j,b}(t)$ and the firms' cash $C_i(t)$.
If deposit insurance is in place we set $w_i = 0$, given that the accounts or the firm's cash is smaller than $\zeta \max\left[PA_{j,b}(t),C_i(t)\right]$, with the deposit insurance parameter $0\leq \zeta < 1$.
Then a one-time bank deposit levy $w_i (M_b+m)$ is transferred from the households and firms to the bailed out bank, where $m$ is an overhead to ensure that the bank has enough equity to resume operations after the bail-out.
The previous owners of $b$ lose their investment and firms and households receive an ownership share of $w_i$ percent of bank $b$.
The flows involved in a bail-out are summarized in figure \ref{Model}C.
\item[Bail-in] A bail-in implements a debt-to-equity conversion.
We assume that the resolution authority seeks to protect depositors as far as possible.
Therefore an amount $M_b+m$ from the claims from other banks are converted into bank $b$'s equity first, where again $m$ is an overhead to ensure that the $b$ can continue its operations after the bail-in.
Only if these funds do not suffice we impose a one-time bank deposit levy following the rules specified under bail-out, but with $w_i$ being non-zero only if the levied deposit is with bank $b$.
In return, ownership rights are transferred to the agents which bailed in the troubled bank, with the prospect of future dividend payments.
The flows of funds involved in a bail-in are shown in figure \ref{Model}D.
\end{description}

\section{Results}

\begin{figure*}
\begin{center}
 \includegraphics[width=140mm]{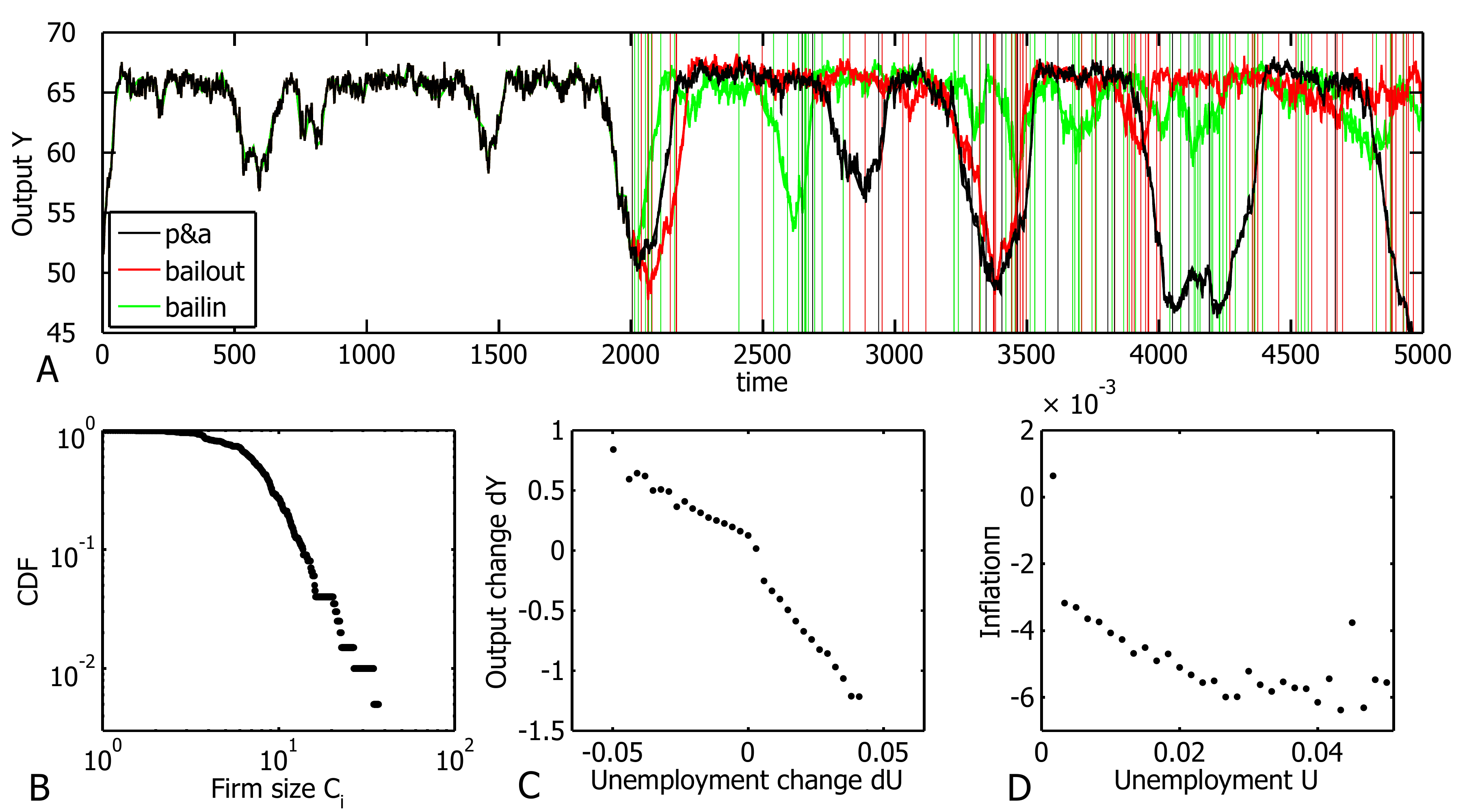}
\end{center}
 \caption{(A) Output $Y$ for a generic run of the model dynamics for the three scenarios 'purchase \& assumption' ('p\&a', black), 'bail-out' (red), and 'bail-in' (green). Each run starts with the same initial configuration and random number seed. Vertical lines indicate that at least one bank's equity become negative at this time. The model produces stylized empirical facts of real economies, such as the (B) firm size distribution, (C) Okun's law, and (D) Philipp's curve.}
 \label{valid}
\end{figure*} 

A MatLab implementation of the Mark I CRISIS model is used and extended by the described crisis resolution mechanisms.
Some of the model parameters are kept fixed, as listed in supporting table S\ref{FixedPar}.
To estimate the robustness of the results we are interested in two different settings (1 and 2) of the remaining parameters listed in supporting table S\ref{VarPar}.

Figure \ref{valid} gives an overview of the model dynamics and shows that stylized empirical facts found in real economies are reproduced (see below).
Simulations were performed with parameter setting 1 and a start equity for banks of $E_b(0)=50$.
Figure \ref{valid}A shows the Output $Y(t)$, which represents the combined value of all produced goods per time step as a function of time.
Results are shown for the three different crisis resolution methods (black for P\&A, red for bail-out, and green for bail-in), starting with the same initial random seed.
At about $t=2000$ the first bank insolvency takes place, and the outputs start to deviate from each other.
Times where at least one bank needs is financially distressed are highlighted by vertical lines in the respective colors.
It becomes immediately apparent that details of the model dynamics are sensitive to the choice of the resolution method, and that crises occur in bursts.
Time periods with a large number of bankruptcies alternate with phases of stability.

The emergent firm sizes $C_i$ are measured as total asset value, unemployment $U(t)$ as the percentage of unemployed workers in the given time step, and inflation $\pi(t)$ is given by the change in retail price index $\pi(t) \equiv \tfrac{\bar p(t)}{\bar p(t-1)}-1$.
Values have been taken in the time period before the first banking crisis occurs.
Figures \ref{valid}B-D show the reproduction of several phenomenological laws found in real economies.
Figure \ref{valid}B confirms a well known empirical Zipf law in the distribution of firm sizes \cite{Axtell01}.
Figure \ref{valid}C shows Okun's law \cite{Prachowny93} for the inverse relation between unemployment change $dU(t) = U(t)-U(t-1)$, and productivity change $dY(t) = Y(t)-Y(t-1)$.
Finally, figure \ref{valid}D shows the Philipp's curve \cite{Blanchard00} as an inverse relationship between unemployment $U$ and inflation $\pi$.

\begin{figure*}
\begin{center}
 \includegraphics[width=160mm]{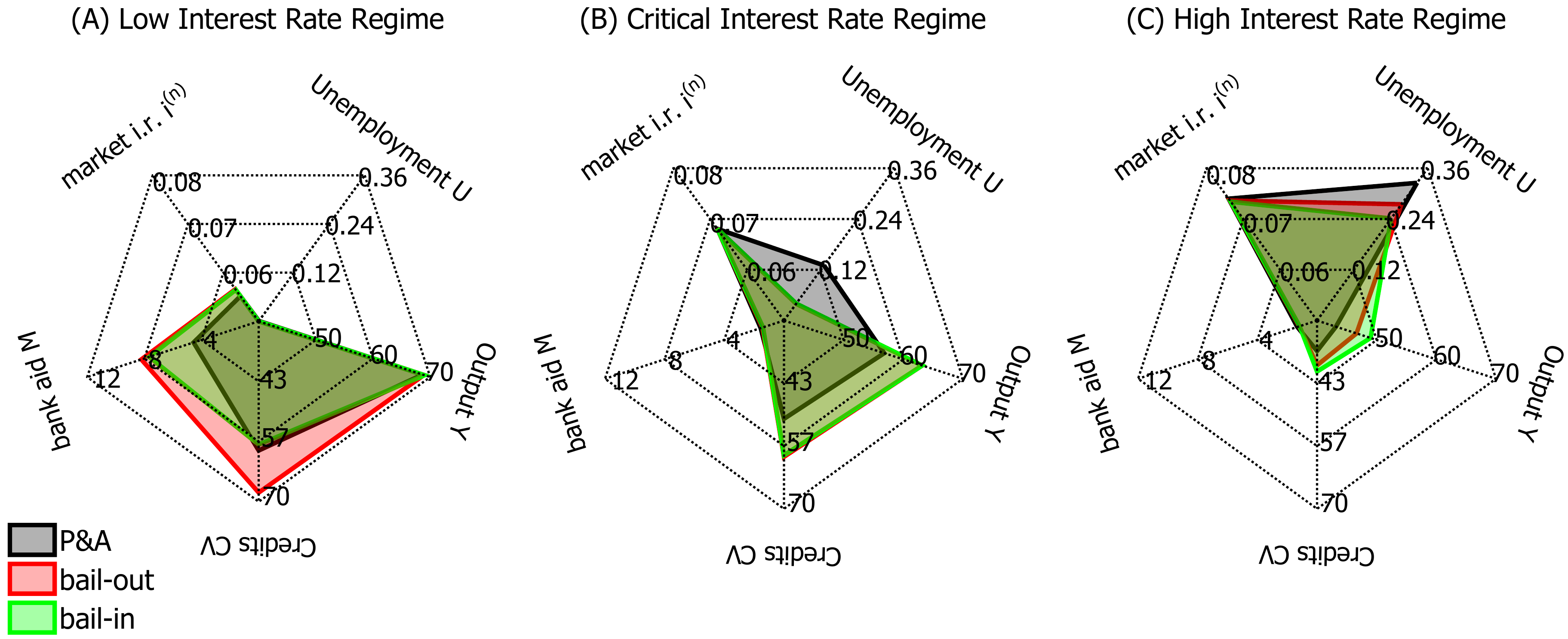}
\end{center}
 \caption{The effects of the P\&A scenario (black), bail-out(red) and bail-in (green) are shown for unemployment $U$, output $Y$, credit volume $CV$, market interest rate $i^{(n)}$, and bank aid $M$ for parameter setting 1 and three different interest rate regimes. (A) In the low interest rate regime ($r_0=0.021$) there is practically zero unemployment, high economic output, and much higher costs associated with bail-outs and bail-ins as compared to the P\&A case. (B) In the critical interest rate regime ($r_0=0.027$) the economy transitions from a healthy state of low unemployment to an unhealthy state of high unemployment and recession. This transition sets in at lower interest rates in the P\&A case compared to bail-outs and bail-ins. (C) The high interest rate regime ($r_0=0.03$) is characterized by low economic productivity at high unemployment. The bail-in crisis resolution mechanism performs consistently better in term of output than its alternatives.}
 \label{spider1}
\end{figure*}

A central and extensively studied feature of the real economy ABM employed in the CRISIS macro-financial model is the existence of a first order phase transition between economic states of low and high unemployment \cite{Gualdi13}.
This phase transition is closely related to the interest rates in the model and, through that, to the refinancing rate $r_0$.
The interest rates for loans offered to firms are roughly given by $r_0$ plus a risk-dependent markup.
A discontinuity appears in the model once this rate approaches the firms' credit contraction threshold, leading to the phase transition \cite{Gualdi13}.
We will therefore compare the results for the different crisis resolution mechanisms in three different interest rate regimes, the low interest rate regime, $i^{(n)}(t) < r^{max}+\pi(t)$, the critical regime, $i^{(n)}(t) \approx r^{max}+\pi(t)$, and the high interest rate regime, $i^{(n)}(t) > r^{max}+\pi(t)$.
Results are discussed for the unemployment $U$, output $Y$, the combined outstanding credit volume of firms $CV$, the interest rate $i^{(n)}$, and the average negative equity in banking crises, $M = \langle M_b \rangle_{E_b(t)<0}$.
Here, $\langle \cdot \rangle_{E_b(t)<0}$ denotes the average over all occurrences of a negative bank equity and hence the application of one of the crisis resolution mechanisms.
Let $t_f$ be the time-step where only one bank is left in the P\&A scenario.
Values have been averaged over 50 iterations and the time-span starting from the first banking crisis until $t=\textrm {min}(t_f,1000)$.
We can now focus on the question of which crisis resolution mechanism shows the best performance in terms of highest economic output, lowest unemployment, and highest financial stability as measured by funds required to save distressed financial institutions.
Results for $U$, $Y$, $CV$, $i^{(n)}$, and $M$ for the generic cases of $r_0$ being below, at, or above the phase transition are shown in figure \ref{spider1}A-C, respectively, for parameter setting 1.
In figure S\ref{spider2} the case for setting 2 is shown.
Figure \ref{spider1} shows a spider plot for each interest rate regime where each crisis resolution mechanism is displayed as a patch (black for P\&A, red for bail-out, green for bail-in) and the area of the patch is proportional to the values of $U$, $Y$, $CV$, $i^{(n)}$, and $M$.
Results for these parameters as a function of $r_0$ are summarized in figure \ref{results1} for parameter setting 1, and for parameter setting 2 in supporting figure S\ref{results2}.
The findings are similar for both sets of parameters.
The left columns show $U$, $Y$, $CV$, $i^{(n)}$, and $M$ as functions of the refinancing rate $r_0$ for the P\&A scenario (black), bail-out (red), and bail-in (green).
The right columns in figures S\ref{results1} and S\ref{results2} show the relative differences between the crisis resolution mechanisms.
If $Z^{(c)}$ is any observable (such as $U$ or $Y$) under crisis resolution mechanism $c=1,2,3$, then $\Delta Z^{(c)}$ is given by $\Delta Z^{(c)} = Z^{(c)} - \langle Z^{(c)} \rangle_c$, i.e. the difference between the observed values and the values averaged over all crisis resolutions $c$.

{\bf Low interest rate regime.}
For both parameter settings the low interest rate regime can be found for $r_0<0.025$.
Results are shown in figures \ref{spider1}A and S\ref{spider2}A for $r_0 = 0.021$.
Since $r_0$ is below its critical value unemployment $U$ is practically zero under each crisis resolution mechanism and values for the output $Y$ are identical; the economy is 'healthy'.
However, the bail-out case shows a slightly higher credit volume $CV$ than the bail-in or P\&A cases.
Since the firms share a higher burden of the financial losses in the bail-out case than in the other two cases, they compensate for this by requesting higher loans from the banks.
The market interest rates are higher in the bail-out and bail-in case compared to the P\&A scenario -- suggesting higher risk-dependent markups in the interest rates offered to firms.
A drastic difference can be seen in the negative equities accumulated by the banks under the three crisis resolution mechanisms.
The value of $M$ is much lower in the P\&A case compared to the bail-out and bail-in scenario.
This suggests that as banks become larger by purchasing and assuming assets of other banks, the risk for further financial losses in the low interest rate regime is greatly reduced.
This is seen in equation \ref{interestrate}, where low interests imply a low number of firm defaults which can be more effectively absorbed by larger banks.

{\bf Critical interest rate regime.}
The critical interest rate regime is found for $0.025<r_0\leq0.027$, see figures \ref{spider1}B and S\ref{spider2}B for $r_0 = 0.027$.
Here the onset of the phase transition can be observed, i.e. unemployment becomes non-zero output decreases.
In the bail-out and bail-in case this onset is at substantially higher values of $r_0$ than for the P\&A scenario.
All parameter values are nearly identical for bail-outs and bail-ins, the output $Y$ is higher compared to the P\&A case due to the later onset of the phase transition.
In this regime the households and firms have still enough wealth to bail-out (or bail-in) banks without leading to a decrease in economic activity.
However, if banks default in this regime and the financial sector has to absorb these losses (as in the P\&A case), this reduces the credit supply compared to the bail-out and bail-in case.
Consequently, there is less economic activity and smaller output $Y$ in the P\&A case.

{\bf High interest rate regime.}
The high interest rate regime is given by $r_0>0.027$ for both parameter settings, see figures \ref{spider1}C and S\ref{spider2}C for $r_0 = 0.03$.
Unemployment soars far above 20\% and output drastically declines; the economy has transitioned into an 'unhealthy' state, a recession.
At the point $r_0=0.027$ two crossovers occur.
Output and unemployment for bail-in and bail-out begin to diverge, with bail-in having higher output at lower unemployment, and the nominal interest rates approach the same value for all three resolution mechanisms.
For higher values of $r_0$ the bail-in strategy consistently outperforms both bail-out and P\&A in terms of output $Y$.
The credit supply in the P\&A case is even smaller than in the critical interest rate regime.
The costs $M$ are comparable in both scenarios.
At the crossover at $r_0=0.027$  the financial burden on households and firms reaches a point where economic activity is impaired stronger in the bail-out case than for bail-ins.
In the high interest rate regime firms and households have not enough savings to assist distressed banks without causing a substantial decline in consumption or production.
Since in the bail-in case only a small set of firms and households need to absorb the financial losses, the remaining parts of the economy perform better than in the bail-out case.

\section{Discussion}

In this work we quantified the economic performance of three different crisis resolution mechanisms which played a pivotal role in the resolution of banking crises in the recent past.
The main difference between these mechanisms is whether the troubled bank is closed down or not, and which sector of the economy has to bear the major burden of the financial losses inflicted by the banking crisis.
Under the P\&A resolution mechanism the distressed bank is closed and its assets and liabilities are purchased and assumed by healthy banks, which thereby pay for the losses.
In a bail-out the financial burden is carried by the taxpayers, in the bail-in the burden is distributed among the troubled banks' debtors.
We tested the performance of these resolution mechanisms within the framework of the Mark I CRISIS model, an ABM equipped with coupled economic and financial sectors.

The performance of the resolution mechanisms is closely related to the state of the model economy, i.e. whether interest rates are high or low.
We found that the P\&A mechanism performs best in a regime where unemployment is low and economic output high, i.e. the economy is 'healthy'.
While in terms of economic productivity there is almost no difference between the different resolution mechanisms, the orderly liquidation of the distressed bank leads to substantially smaller losses to other banks and therefore increased financial stability compared to bail-outs or bail-ins.
This is because the P\&A case resolution mechanism leads to larger banks which can absorb the -- in the low interest rate regime -- comparably small losses due to firm defaults much more effectively than banks which are drip-fed by bail-ins or bail-outs.
However, in the recession regime, characterized by high unemployment and comparably low economic productivity, this does not hold.
In this regime the bail-in mechanism outperforms both the P\&A and the bail-out mechanism in terms of higher economic output and stability.
Here the banking sector alone can no longer cope with the increasing number of firm defaults and credit supply dwindles under the P\&A mechanism.
In the bail-in case the financial losses are absorbed by, and confined to a smaller part of the economy, the remaining parts of the economy perform better than in the bail-out case.
In the intermediate economic state between the healthy and unhealthy, bail-outs and bail-ins lead to almost the same results and both outperform the P\&A mechanism.
In this case the households and firms are wealthy enough to allocate funds to troubled financial institutions without reducing productivity and consumption.

A number of questions and further research agendas immediately arise as a consequence of this work.
First, it remains to be seen how the results depend on the actually imposed liquidation procedure.
In the P\&A case the number of banks decreases over time and we stop the simulation if only one bank is left.
It is interesting to see what happens if we re-populate the financial sector with new banks?
The evolutionary perspective of having competing banks with different strategies or even regulatory regimes also remains to be explored.
Further, it would be interesting to investigate a mix of resolution mechanisms.
For example, banks could be only partially liquidated or only if certain conditions are met, such as the bank being small enough.
How do our results depend on the choice of the economic production model?
What if, instead of the Mark 1 economic model, firms are endowed with capital and labor and produce goods according to a Cobb-Douglas function or, say, the $AK$ growth model?
Does this impact the performance of the resolution mechanisms?
It will be possible to study these and related questions in the future where the CRISIS framework, which provides a suite of economic and financial model building-blocks which can be plugged together and extended.
So far, our main results are obtained within the cosmos of the Mark 1 CRISIS model:
We showed that (i) it is beneficial to let distressed banks default only if the overall state of the economy is healthy enough and (ii)
there are no economic conditions under which a taxpayer-funded bail-out outperformed the bail-in mechanism with private sector involvement.

\section{Acknowledgments}
We acknowledge financial support from FP7 projects CRISIS agreement no. 288501, LASAGNE, agreement no. 318132, and MULTIPLEX, agreement no. 317532.

\clearpage

\onecolumngrid
\appendix

\setcounter{figure}{0}

\large 
\parbox[c]{160mm}{Supporting Information for 'To bail-out or to bail-in? Answers from an agent-based model'}

\normalsize

\begin{table*}[h]
\caption{Overview and description of fixed model parameter.}
\begin{tabular}{llr}
Symbol & Parameter & Value \\
\hline
$B$ & number of banks & 20 \\
$I$ & number of firms & 100 \\
$J$ & number of worker & 700 \\
$E_b(0)$ & banks' start equity & 10 \\
$\alpha$ & labor productivity & 0.1 \\
$w$ & wage & 1\\
$z$ & number of applications in consumption good or credit market & 2 \\
$\kappa^{min}$ & minimal cash reserve ratio & 0.005 \\
$\lambda^{max}$ & maximal leverage ratio & 100 \\
$\tau$ & debt reimbursement rate & 0.05 \\
$\mu$ & interest rate coefficient & 0.2 \\
$r^{max}$ & credit contraction threshold & 0.05 \\
$r^{(ib)}$ & interbank loan interest rate & 0
\end{tabular}
\label{FixedPar}
\end{table*}

\begin{table*}[h]
\caption{Overview and description of variable model parameter.}
\begin{tabular}{llrr}
Symbol & Parameter & Setting 1 & Setting 2 \\
\hline
$\delta$ & dividends & 0.25 & 0.5 \\
$\phi$ & credit contraction parameter & 0.75 & 0.8 \\
$c$ & propensity to consume & 0.85 & 0.8 \\
$m$ & crisis resolution overhead & 1 & 0.5 \\
$\zeta$ & deposit insurance & 0.0 & 0.05
\end{tabular}
\label{VarPar}
\end{table*}

\begin{figure*}[h]
\begin{center}
 \includegraphics[width=160mm]{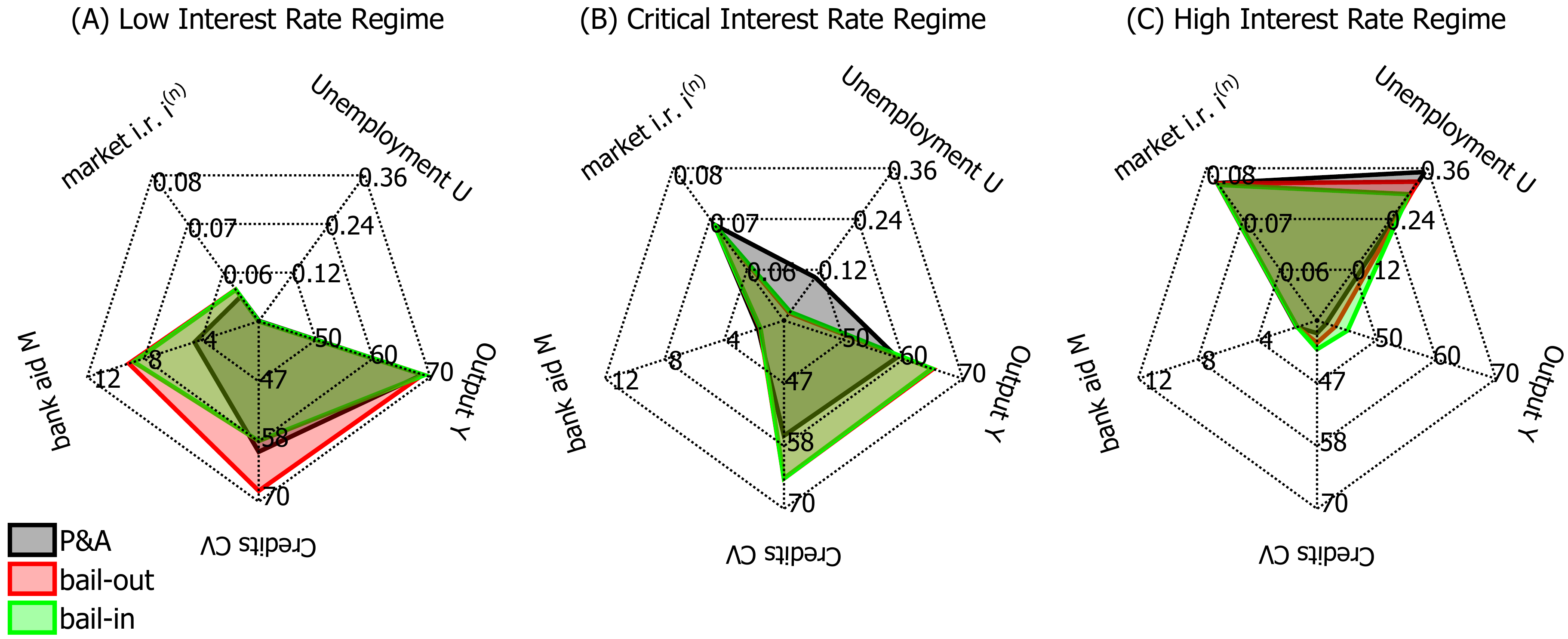}
\end{center}
 \caption{Same as figure \ref{spider1} for parameter setting 2.}
 \label{spider2}
\end{figure*}

\begin{figure*}[h]
\begin{center}
 \includegraphics[width=120mm]{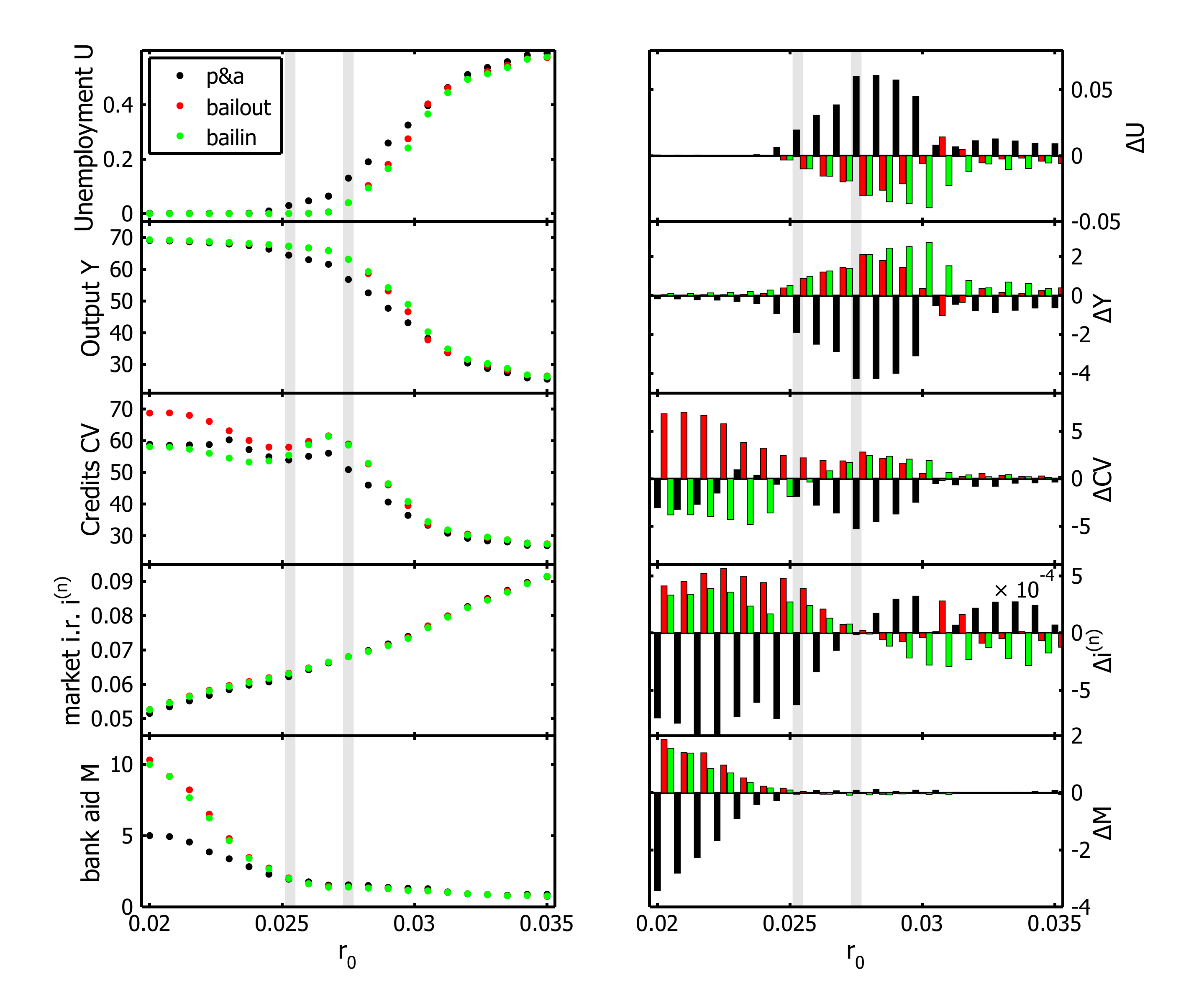}
\end{center}
 \caption{Effects of the P\&A scenario (black), bail-out(red) and bail-in (green) are shown for (left column) unemployment $U$, output $Y$, credit volume $CV$, market interest rate $i_n$, and bank aid $M$ as functions of $r_0$ for parameter setting 1. The right column shows the relative differences across different crisis resolution mechanisms for these parameters. In the low interest rate regime ($r_0<0.025$, indicated by the first vertical gray patch) there is practically zero unemployment, high economic output, and much higher costs associated with bail-outs and bail-ins as compared to the P\&A case. In the critical interest rate regime $0.025<r_0<0.027$ the economy transitions from a healthy state of low unemployment to an unhealthy state of high unemployment. This transition sets in at lower interest rates in the P\&A case compared to bail-outs and bail-ins. The high interest rate regime ($r_0>0.027$, given by the second vertical gray patch) is characterized by low economic productivity at high unemployment. The bail-in crisis resolution mechanism performs consistently better in term of output than its alternatives.}
 \label{results1}
\end{figure*} 

\begin{figure*}[h]
\begin{center}
 \includegraphics[width=120mm]{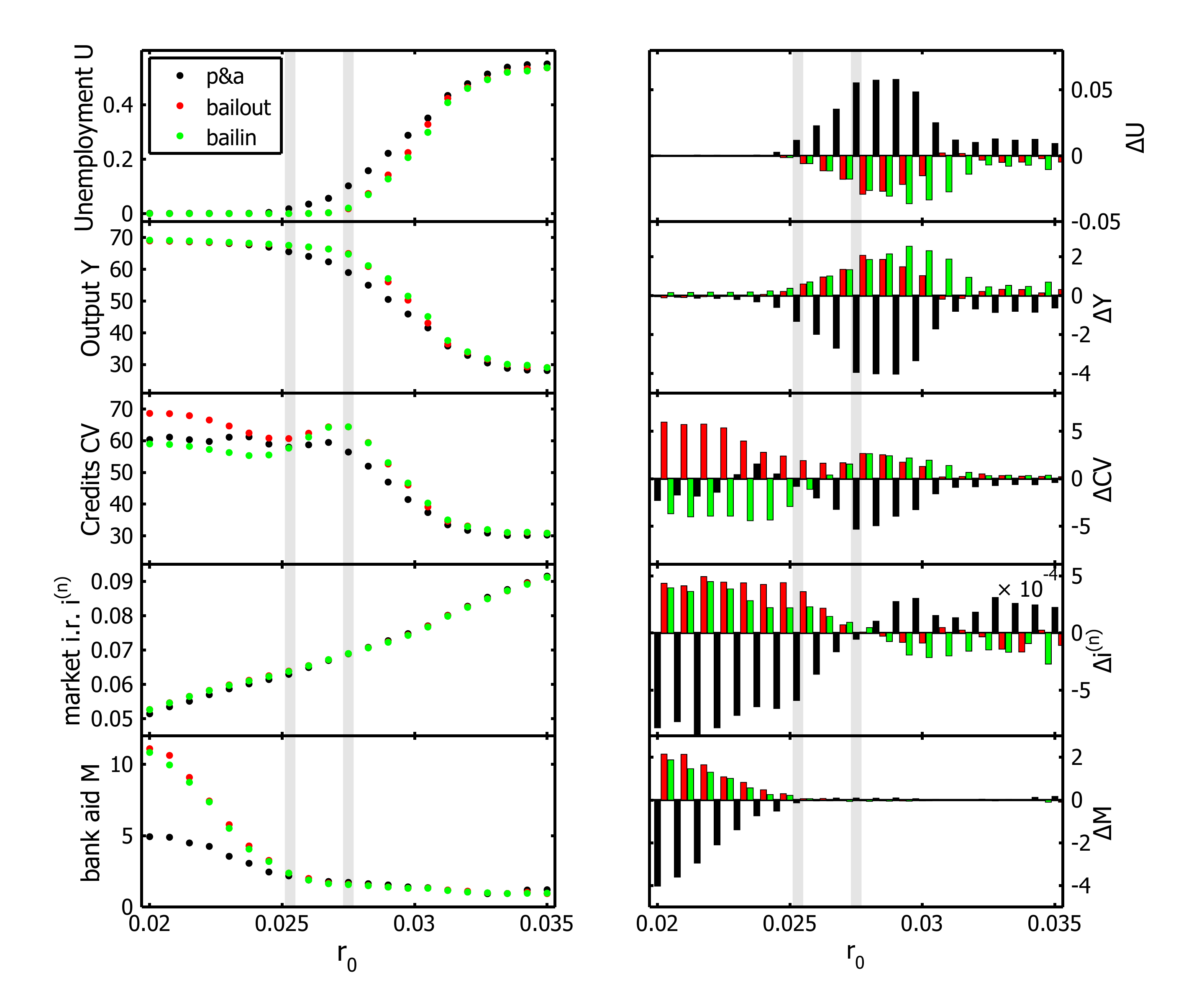}
\end{center}
 \caption{Same as figure \ref{results1} for parameter setting 2. All observations and findings from setting 1 apply here too.}
 \label{results2}
\end{figure*}


\begin{thebibliography}{}

\bibitem{Stavarek13} D. Stav\'arek (2013), Lessons learned from the 2013 banking crisis in Cyprus. {\it European Financial Systems 2013}, 312.

\bibitem{Wilson12} J. Wilson (2012), Cyprus requests eurozone bailout. {\it Financial Times}, retrieved 25/07/2012.

\bibitem{AlJazeera13} Al Jazeera (2013), Cyprus asks EU for financial bailout, retrieved 16/01/2013.

\bibitem{Guardian13} The Guardian (2013), Cyprus eurozone bailout prompts anger as savers hand over possible 10\% levy, retrieved 16/03/2013.

\bibitem{Otker11} I. \"Otker-Robe, A. Narain, A. Ilyina, J. Surti (2011), The too-important-to-fail conundrum: impossible to ignore and difficult to resolve. {\it IMF Staff Discussion Note} SDN/11/02.

\bibitem{Zhou12} J. Zhou, V. Rutledge, W. Bossu, M. Dobler, N. Jassaud, M. Moore (2012), From bail-out to bail-in: mandatory debt restructuring of systemic financial institutions. {\it IMF Staff Discussion Note} SDN/12/03.

\bibitem{DeGrauwe13} P. DeGrauwe (2013), The new bailin doctrine: a recipe for banking crises and depression in the eurozone. {\it CEPS Commentary}, retrieved 04/04/2013.

\bibitem{Reuters13} Reuters (2013), Bank bond investors should fear fast-tracked EU bail-in, retrieved 12/12/2013.

\bibitem{Stern04} G.H. Stern, R.J. Feldman, {\it Too big too fail: the hazards of bank bailouts} (Brookings Institution Press, 2004).

\bibitem{GAO12} US Government Accountability Office (2012), Capital Purchase Program: revenues have exceeded investments, but concerns about outstanding investments remain. GAO-12-301.

\bibitem{EFSF13} EFSF (2013), FAQ about European Financial Stability Facility and the new ESM, 21/01/2013 (retrieved 13/02/2014).

\bibitem{EC13} European Commission (2013), Balance of payments - European Commission. ec.europa.eu, retrieved 13/02/2014.

\bibitem{Kiyotaki97} N. Kiyotaki, J. Moore (1997), Credit cycles. {\it Journal of Political Economy} {\bf 105}(2), 211-48.

\bibitem{Bernanke99} B. Bernanke, M. Gertler, S. Gilchrist, The financial accelerator in a quantitative business cycle framework. In: {\it Handbook of Macroeconomics, Volume 1}, ed. J.B. Tayler, M. Woodford (Elsevier, 1999).

\bibitem{Dubey05} P. Dubey, J. Geanakoplos, M. Shubik (2005), Default and punishment in general equilibrium. {\it Econometrica} {\bf 73}(1), 1-37.

\bibitem{Tsomocos03} D. P.Tsomocos (2003), Equilibrium analysis, banking and financial stability. {\it Journal of Mathematicel Economics} {\bf 39}, 619-55.

\bibitem{Goodhart06} C. A.E. Goodhart, P. Sunirand, D. P. Tsomocos (2006), A model to analyse financial fragility. {\it Economic Theory} {\bf 27}, 107-42.

\bibitem{DeWalque10} G. De Walque, O. Pierrard, A. Rouabah (2010), Financial (in)stability, supervision and liquidity injections: A dynamic general equilibrium approach. {\it The Economic Journal} {\bf 120}(549), 1234-61. 

\bibitem{Goodhart12} C. A.E. Goodhart, A. K. Kashyap, D.P. Tsomocos, A. P. Vardoulakis (2012), Financial regulation in general equilibrium. {\it NBER Working Paper} No. 17909.

\bibitem{McGuire12} C.L. McGuire {\it Simple tools to assist in the resolution of troubled banks} (World Bank, 2012).

\bibitem{Thurner13} S. Thurner, S. Poledna (2013), DebtRank-transparency: Controlling systemic risk in financial networks. {\it Scientific Reports} {\bf 3}, 1888.

\bibitem{Poledna14} S. Poledna, S. Thurner (2014), Elimination of systemic risk in financial networks by means of a systemic risk transaction tax, arXiv.org: 1401.8026v2.


\bibitem{Boss05} M. Boss, M. Summer, S. Thurner (2005), The network topology of the interbank market. {\it Quantitative Finance} {\bf 4}, 677-84.

\bibitem{Iori08} G. Iori, G. De Masi, O.V. Precup, G. Gabbi, G. Caldarelli (2008) A network analysis of the Italian overnight money market, {\it Journal of Economic Dynamics and Control} {\bf 32}, 259-78.

\bibitem{Cacciolo12} F. Caccioli, J.-P- Boucaud, J. Farmer (2012), Impact-adjusted valuation and the criticality of leverage, arXiv.org: 1204.0922. 

\bibitem{Battiston12} S. Battiston, M. Puliga, R. Kaushik, P. Tasca, G. Caldarelli (2012), Too central to fail? Financial networks, the FED and systemic risk. {\it Scientific Reports} {\bf 2}.

\bibitem{Thurner12} S. Thurner, J. Farmer, J. Geanakoplos (2012), Leverage causes fat tails and clustered volatility, {\it Quantitative Finance} {\bf 12}, 697-707.

\bibitem{Poledna12} S. Poledna, S. Thurner, J. Farmer, J. Geanakoplos (2014), Leverage-induces systemic risk under Basle II and other credit risk policies, arXiv.org: 1301.6114.

\bibitem{DelliGatti08} D. Delli Gatti, S. Desiderio, E. Gaffeo, P. Cirillo, M. Gallegati, {\it Emergent macroeconomics: An agent-based approach to business fluctuations} (New Economic Windows, Spriger, 2008).

\bibitem{Gaffeo08} E. Gaffeo, D. Delli Gatti, S. Desiderio, M. Gallegati (2008), Adaptive microfoundations for emergent macroeconomics. {\it Technical Report} 0802.

\bibitem{DelliGatti11} D. Delli Gatti, S. Desiderio, E. Gaffeo, P. Cirillo, M. Gallegati, {\it Macroeconomics from the bottom-up} (Springer Milan, 2011).

\bibitem{Schumpeter42} J. A. Schumpeter, {\it Capitalism, socialism and democracy} (Harper, NY, 1942).

\bibitem{Klimek12} P. Klimek, R. Hausmann, S. Thurner (2012), Empirical confirmation of creative destruction from world trade data. {\it PLOS ONE} {\bf 7}(6): e38924.

\bibitem{FDIC03} Federal Deposit Insurance Corporation, {\it Resolutions Handbook} (FDIC, 2003).

\bibitem{Wright09} R. E. Wright, {\it Bailouts: Public Money, Private Profit} (Columbia University Press, New York, 2009).

\bibitem{FDIC12} Federal Deposit Insurance Corporation, Bank of England, {\it Resolving globally active, systemically important, financial institutions} (FDIC, BoE, 2012).

\bibitem{Bryant93}
R.C. Bryant, P. Hooper, C. Mann (1993), Evaluating policy regimes: new research in empirical macroeconomics (Brookings, Washington D.C.).

\bibitem{Axtell01} R. Axtell (2001), Zipf distribution of US firm sizes. {\it Science} {\bf 293} (5536), 1818-20.

\bibitem{Prachowny93} M.F.J. Prachowny (1993), Okun's Law: Theoretical foundations and revised estimates. {\it The Review of Economics and Statistics} {\bf 65} (2), 331-6.

\bibitem{Blanchard00} O. Blanchard (2000), Macroeconomics (Second ed.). Prentice Hall.

\bibitem{Gualdi13} S. Gualdo, M. Tarzua, F. Zamponi, J.-P. Bouchaud (2013), Tipping points in macroeconomic agent-based models, arXiv.org: 1307.5319.


\end{thebibliography}
\end{document}